\documentclass[preprint]{aastex} 

\begin{document}

\title{ON THE INTERACTIVE-BEATING-MODES MODEL: GENERATION OF ASYMMETRIC MULTIPLET STRUCTURES AND EXPLANATION OF THE BLAZHKO EFFECT}

\author{Paul H. Bryant}
\affil{BioCircuits Institute (formerly Institute for Nonlinear Science), \\University of California, San Diego, La Jolla, CA 92093, USA}
\email{pbryant@ucsd.edu}

\begin{abstract}
This paper considers a nonlinear coupling between a radial and a nonradial mode of nearly the same frequency.  The results may be of general interest, but in particular have application to the ``beating-modes model" of the Blazhko effect which was recently shown to accurately reproduce the light curve of RR Lyr.  For weak coupling, the two modes do not phase-lock and they retain separate frequencies, but the coupling nevertheless has important consequences.  Upon increasing the coupling strength from zero, an additional side-peak emerges in the spectrum forming an asymmetric triplet centered on the fundamental.  As the coupling is further increased, the amplitude of this side-peak increases and the three peaks are also pulled towards each other, decreasing the Blazhko frequency.  Beyond a critical coupling strength, phase-locking occurs between the modes.  With appropriate choice of coupling strength, this ``interactive beating-modes model" can match the side-peak amplitude ratio of any star.  The effects of nonlinear damping are also explored and found to generate additional side-peaks of odd order.  Consistent with this, the odd side-peaks are found to be favored in V808 Cyg.  It is also shown that the Blazhko effect generates a fluctuating ``environment" that can have a modulatory effect on other excited modes of the star.  An example is found in V808 Cyg where the modulation is at double the Blazhko frequency.  An explanation is found for this mysterious doubling, providing additional evidence in favor of the model.
\end{abstract}


\section{INTRODUCTION}
\label{introduction}
The Blazhko effect is a modulation of the pulsations found in some variable stars of the RR Lyrae type.  The origin of the effect has been the subject of debate ever since it was first reported in 1907 by Russian astronomer Sergey Nikolaevich Blazhko \citep{Blazhko}.  See the recent reviews by \citet{Kovacs2009}, \citet{Kolenberg2012}.   Various models have been proposed over the years to explain the effect including the ``beating-modes model" which was recently presented by the author \citep{Bryant2015}.  Other recent models can be found and discussed in Section 4 of \citet{Bryant2015}. In a recent paper, \cite{Chadid2014} hypothesize that the origin of the Blazhko effect is a dynamical interaction between a multi-shock structure and an outflowing wind in a coronal structure.

The ``beating-modes model" proposes that the modulation is not a true modulation at all, but rather a beat-frequency between two vibrational modes with slightly different frequencies, one of which is highly non-sinusoidal.  That model achieves a good fit with Kepler data for RR Lyr and, more importantly, reproduces key features of the observed light curve including the pulsation waveform, the upper and lower Blazhko envelope functions which are offset from each other in phase, and the motion, disappearance, and reappearance of the so called ``bump" that appears on the pulsation waveform.  The basic idea for the model is not new \citep[see, e.g.][]{Cox1993}, but previous efforts made no attempt to show that the model could achieve a good fit to actual data, and it has been given little, if any, serious consideration.  Other evidence in its favor include an ability to explain the mysterious double-maxima waveform found in the star V445 Lyr \citep[Section 3.1 in][]{Bryant2015} and also, as is shown in Section~\ref{modulation} below, the ability to explain a double-Blazhko modulation observed in V808 Cyg.

This paper examines an important modification of the beating-modes model in which the two modes may interact through nonlinear coupling.  This will be referred to as the ``interactive-beating-modes model".  In Section~\ref{analytic} we analyze the interaction in the adiabatic limit.  We consider the case where one of the modes (presumably the fundamental) is much more robust than the other and is consequently affected much less by the interaction.  The lowest order nonlinear coupling is considered, and the dynamics are found to generate an opposing side-peak, producing a triplet spectrum with asymmetry dependent on the coupling strength.  In the adiabatic limit, this is a pure triplet with no additional side-peaks.  This model can match any possible side-peak amplitude ratio.  That ratio is known to vary widely in observational data.  RR Lyr, for example, is known to have a very asymmetric triplet spectrum with ratio measured using Period04 software to be 0.194 for second quarter \emph{Kepler} data with the larger side-peak on the right of the fundamental,  while the ratio for V808 Cyg was measured to be 0.857 with the larger side-peak on the left.  Although the idea that the opposing side-peak is generated by a nonlinear mixing process is implicit in the beating-modes model, it was explicitly discussed in a general way in \citet{Breger2006}.

In Section~\ref{dissipative} nonlinear dissipative effects are considered, and it is found that additional side-peaks of odd order emerge in the spectrum.  Consistent with this, data from V808 Cyg is shown to favor third order over second order side-peaks of the fundamental.  (The presence of the weaker second order peaks can also be accounted for.)

Modes with extremely small detuning from the fundamental become phase-locked to it and effectively disappear from the spectrum.  Assuming that the phase-locked state is stable (and this is generally the case) these modes \emph{will not} produce a Blazhko effect.  On the other hand, when the modes are not phase-locked, the relative phase continues to increase by $2\pi$ every Blazhko period.  The cases that produce the most dramatic Blazhko effect are those for which the detuning is just slightly too large to allow phase-locking to occur.  The interaction is found to pull the frequency of the nonradial mode toward the fundamental, substantially reducing Blazhko frequency from the value of the detuning in the absence of coupling.  Note that this model is quite different from one claiming that the Blazhko effect comes from a dynamical oscillation between the fundamental and a mode that is phase-locked to it.  ``Phase-locked" in that case does not mean fixed relative phase, but rather a relative phase that oscillates within a range of values and does not increase without bound. \citet{Buchler2011} found such an oscillatory case for a model involving the ninth overtone (oscillatory cases 2a and 2b rather than the fixed point case 1c on their page 2).  \citet{Nowakowski2001} tried to find such an oscillatory phase-locked solution between the fundamental and a near-resonant nonradial mode and were unable to do so, suggesting that this is not a viable mechanism for the Blazhko effect.

There appears to be good observational evidence of nonradial modes in RR Lyrae stars.  \cite{Moskalik2003}, in a study of multiperiodic RR Lyrae variables of the Galactic Bulge, find 25 stars of the RR0-$\nu$1 type which are fundamental mode pulsators that have a near-resonant peak in the spectrum forming a doublet, with no opposing side-peak (at least not visible above the noise).  Since these peaks seem too close together to be generated by two different radial modes, this would seem to be good evidence that the smaller of the peaks was generated by a nonradial mode.  The only alternative is to imagine that the single side-peak is generated by modulating the fundamental in a manor that is special in the sense that it suppresses all side-peaks except the one observed.  \citet{Moskalik2003} state ``It is not clear at present if so strongly asymmetric triplets (or pure doublets) can be reproduced by the resonant mode coupling theory".  \cite{Chadid1999} find a multiplet structure in variations of the line profile of RR Lyr which they suggest may be indicative of the presence of a nonradial mode.

There is also observational evidence for nonradial modes that are not near-resonant with the fundamental or with the first radial overtone.  In the spectrum of V445 Lyr, \cite{Guggenberger2012} suggest that the peak designated $f_N$ is likely nonradial.  In the spectrum of V1127 Aql, \cite{Chadid2010} suggest that a peak designated $f'$ could be a nonradial p-mode and a peak designated $f''$, found below the fundamental, could be a g-mode.  Another likely example is the doubled peak in V808 Cyg described below in Section~\ref{modulation}.

\cite{VanHoolst1998} in their Figure 1, indicate that nonradial modes very close in frequency to the fundamental may have a growth rate advantage, possibly explaining the tendency for these modes to be found in very close proximity.  Their results suggest the observed modes might be dipolar modes of radial order near 140.  Results of \citet{Cox1993} and \citet{Cox2009} suggest a much lower radial order (4 and 12 respectively).  However, all of these calculations are for a static star, without the presence of the high amplitude fundamental mode.  Because of its extreme nonlinear character, the fundamental would seem likely to have very dramatic effects on the characteristics of the other modes such as their frequencies, growth rates, spatial waveform, etc., which may be very difficult to predict.

In Section~\ref{modulation} it is shown that the model provides an explanation for a peak observed in V808 Cyg that is apparently modulated at twice the Blazhko frequency.  The explanation is based on the effects of the fluctuating environment provided by the Blazhko effect and specifically depends on the involvement of the nonradial mode.  The effect therefore would seem to be incompatible with Blazhko models that assume the fundamental mode is being modulated by some process, which is true for almost all Blazhko models.  There are terms in the amplitude equations that generate these environmental effects, which in some other papers are referred to as ``saturation" terms.  See e.g. Equations~6 and~7 in \citet{Nowakowski2001} in which the ``R" terms are the resonant terms and the ``S" terms are the saturation or environmental terms.  Environmental terms can provide a limiting effect on the growth of a mode, as implied by the word ``saturation", but they can also result in frequency shifts and modulation.  These terms will not result in phase-locking or direct transfer of energy between modes.

\section{ADIABATIC RESULTS}
\label{analytic}
For relatively small excitations from equilibrium the stellar dynamics can be described (in the adiabatic approximation) in terms of a set of normal modes characterized by a set of eigenfunctions and corresponding set of normal mode displacements $x_1, x_2, x_3, ...$ \citep[see e.g.][]{Unno}.  In the treatment presented here, these displacements are real, not complex, and the modal interactions are described by a set of real differential equations.  Amplitude equations in the astrophysical literature are most commonly found in the complex form; the conversion of the results presented here to the complex form is straightforward and given in Appendix~\ref{complex}.  Both the real and complex forms are discussed in \citet{Nayfeh1994}.  The equations for the adiabatic case are generated using a potential energy approach.  This approach does not seem to be discussed much in papers on stellar dynamics, but can be found in other literature relating to normal modes \citep[see e.g.][Section 2, paragraph 1]{King1996}.  One advantage to this approach is to automatically generate equations that satisfy energy conservation.  An arbitrary displacement from equilibrium will result in an increase in the potential energy $V$ of the system which is the combination of the gravitational potential energy and the internal energy.  If this displacement is expanded in terms of the eigenfunctions, then the potential energy can be expressed in terms of the normal mode displacements.  Since the potential is a function of the normal mode displacements, it can be expanded as a power series in them.  The lowest order terms in this power series will be quadratic with no cross terms between them, i.e. no terms such as $x_1x_2$ (the eigenfunctions are chosen such that this is the case).  As a result, in the limit of low level excitations, the normal modes are completely independent harmonic oscillators.  But for any finite level of excitation, there are higher order terms in the potential energy, cubic and above, and these may include cross terms between the normal mode displacements.  This will result both in nonlinearity within the equations of motion for the $x_i$ but will also result in coupling terms between them.  The effective ``force" that appears in the equation of motion of a particular mode is determined by the negative of the derivative of the potential energy with respect to the displacement variable corresponding to that mode.  Each mode will have an effective ``mass" or inertia $M_i$ associated with it.  However, by independently rescaling the amplitudes of each of the modes, the $M_i$ can all be set to unity.  The kinetic energy of mode $i$ can then be expressed as $\dot{x}_i^2/2$ and its equation of motion is given by:
\begin{equation}\label{adiabatic1}
\ddot{x}_i = - \omega_i^2x_i + f_i(\textbf{x}) = -\partial V(\textbf{x}) / \partial x_i
\end{equation}
where $ f_i(\textbf{x})$ is the nonlinear part of the effective force for mode $i$, and $V(\textbf{x})$ is the potential energy, both as nonlinear functions of all the displacements $x_1, x_2, x_3, ...$.  There are terms in the power series expansion of the potential involving only one mode: $B_{i,n} x_i^n$ where $B_{i,n}$ is a coefficient.  The lowest nonzero term in this expansion is the quadratic, and it determines the frequency: $\omega_i^2 = 2B_{i,2}$.  The higher order terms can result in a non-sinusoidal distortion to the oscillation and/or a shift in the frequency.  Most terms in the potential involve two or more modes and can result in coupling between them.  Note that although this approach is most accurate when all nonlinear terms are small, we know that for RR Lyrae and some other types of variable stars the pulsation is often highly non-sinusoidal.  It is not clear to what extent it is useful to try and model this using the $B_{i,n} x_i^n$ terms or some other representation of a nonlinear function of $x_i$.  \citet{Stellingwerf1972} studied a simple ``one zone" nonlinear oscillator model of this type which can generate the sawtooth-like velocity function that is ubiquitous in Cepheid and RR Lyrae stars.  But this kind of approach may be lacking in certain respects due to its simplicity.  In particular, the spacial character of a mode at high amplitude may deviate substantially from its eigenfunction and may be more non-sinusoidal near the surface than internally.  \citet{Nayfeh1994} show how nonlinear normal modes can be treated in a more precise way, while at the same time be parameterized as a single position/velocity pair even though the state for that nonlinear mode is no longer an eigenfunction and can be expanded as a linear combination of the eigenfunctions; see their Equation 12.   We also know from the stellar spectra that one or more low amplitude modes can coexist with the main oscillation \citep[see, e.g.,][]{Guggenberger2012} and so it is reasonable to assume that the system can still be approximately described in terms of nonlinear interactions between this set of modes.  The cross terms can lead to phase-locking if sufficiently strong, or can cause modulation as a result of the slipping relative phase between the modes and corresponding periodic change in the effects of the cross term on each of the modes.  These effects are the most significant when there is a near-resonance.  In this case the corresponding term or terms in the potential will have some component at a very low (or zero) frequency.  When a mode is non-sinusoidal, such as the fundamental in this case, it will have harmonic components and the presence of these additional frequencies can sometimes cause a term in the potential to become near-resonant when it was not previously.  Within the adiabatic approximation, energy lost by one mode must be gained by the modes with which it is interacting through the nonlinear coupling terms.

For the case of interest, there are two modes of nearly the same frequency: the fundamental which is assumed to be very robust and highly non-sinusoidal, and a weaker mode that is nearly sinusoidal and is assumed to be a nonradial mode.  We will refer to the fundamental as mode 0 and the nonradial mode as mode 1, represented respectively by $x_0$ and $x_1$.  The corresponding angular frequencies are $\omega_0$ and $\omega_1$, and it should be noted that these are the frequencies in the absence of nonlinear effects; the peaks observed in the stellar spectra, especially that of mode 1, may be shifted away from these values as will be shown below.  For the sake of convenience, the dominant side-peak in the spectrum may be referred to as $f_1$ even though technically it should be referred to as ``the observed $f_1$".  There is no peak in the spectrum at the actual (unshifted) $f_1$.

The robust character of the fundamental should be reflected in a comparison of their average kinetic energies: $\langle \dot{x}_0^2/2\rangle \gg  \langle\dot{x}_1^2/2 \rangle$.  The character of the nonradial mode as a function of polar and azimuthal angles is derived from a spherical harmonic $Y_l^m$ [see e.g. \citet{Unno}, Equation 13.60 or \citet{Aerts}, Equations 1.1, 1.2 and 1.3].  Because $x_1$ is nonradial, the potential energy $V(\textbf{x})$ can only contain cross-terms with powers of $x_1$ greater than one.  So terms of the form $Cx_0^n x_1$ are not allowed for any $n$.  This is because a radial mode to any power is still an entirely radial function.  So it has no angular dependence, which in terms of spherical harmonics is the constant $Y_0^0$.  This is of course orthogonal to any other $Y_l^m$ including the one corresponding to $x_1$.  The net potential energy for such terms, being determined by an integral over the sphere, will thus necessarily be zero.  On the other hand, putting any spherical harmonic to any even power will produce a function that is everywhere positive or zero.  Expanding this resulting function in terms of spherical harmonics will always produce a $Y_0^0$ component, and this can interact with the radial mode.  This is also true for odd powers of 3 or more, for nonradial modes which have $m=0$ and $l$ even\footnote{\label{sym1}
Some modes have symmetry properties that prevent odd powers from producing a $Y_0^0$ component.  This includes all modes with $m \ne 0$ and also the modes with $m=0$ for which the spherical harmonic is antisymmetric about the equator, i.e. those with $l$ odd.  In the latter case if we make a change of variables $\theta^\prime = \pi - \theta$ we reproduce the exact same mode with inverted amplitude.  Since the radial mode has no $\theta$ dependence, this should mean that terms in the potential involving both $x_0$ and $x_1$ should be invariant under a sign change for $x_1$, e.g. $Cx_1^kx_0^n = C (-x_1)^kx_0^n$.  For $k$ odd this can only be true if $C=0$, i.e., odd powers of $k$ are disallowed.  Similarly, if we consider a mode with $m \ne 0$ we can make a change of variables $\phi^\prime = \phi + \pi/m$ we again have an inverted amplitude and the same argument holds.}.
So the allowed cross terms have the form $C_{jk}x_0^jx_1^k$ where $k \ge 2$ with the additional condition that if $k$ is odd then the nonradial mode must have $m=0$ and $l$ even.  We will mainly be interested in terms (if any) which are near-resonant; the other allowed terms will typically be neglected.

We will initially assume that $m=0$; the case of nonzero $m$ is discussed briefly in Section~\ref{nonzerom}.  Here we will consider only the cases that are quadratic in $x_1$ as these are lowest order and therefore likely to be much stronger than those being neglected.  Two terms in the potential energy that might result in a significant interaction are $C_1 x_0 x_1^2$ and $C_2 x_0^2 x_1^2$.  The $C_1$ term is only relevant if we allow the fundamental to be strongly non-sinusoidal, possibly resulting in $x_0$ having a strong second harmonic component.  This possibility is not generally considered in the astrophysical literature.  However, as is discussed below, the presence or absence of this term has no qualitative effect on the results presented in this paper, as the two terms have similar effects on the dynamics.  The $C_2$ term does not depend on any harmonic content but is a higher order term.  The corresponding force terms that appear in the equation for $\ddot{x}_1$ are $-2C_1 x_0 x_1$ and $- 2C_2 x_0^2 x_1$.  Note that the second of these, if converted to the complex form as described in Appendix~\ref{complex}, will generate terms corresponding to both the second and fourth terms on the right hand side of Equation 7 in \citet{Nowakowski2001} as well as other terms that can be neglected as nonresonant.  The remaining nonlinear terms in their Equation 7 (terms 3, 5 and 6) can be neglected in our application as they are higher order in $x_1$.  (terms 5 and 6 are also disallowed by symmetry for $l$ odd as mentioned in their paper.)  Term 3 in their Equation 7 does play an important role in nonlinear damping as we discuss in Section~\ref{dissipative} and Appendix~\ref{complex}.

The fundamental, due to its higher energy content, will be effected only slightly by its coupling to the nonradial mode.  We will initially make the approximation that we can neglect the effect of $x_1$ on $x_0$ entirely and that therefore $x_0$ can be approximated as a fixed periodic function with frequency $\omega_0$.  (The more general case with separate equation for $\ddot{x}_0$ and dissipation is discussed briefly in Section~\ref{x0x1} and is found to reproduce the results given here.)  If this function has a significant second harmonic component, it will produce a near-resonance in the $C_1$ term.  The the square of this function will generate a second harmonic component that will produce a near-resonance in the $C_2$ term.  These terms have the same $x_1$ dependence so they can be combined together and the time origin can be chosen so that the net second harmonic is a pure cosine with some nonnegative amplitude.  Contributions to this combined term from terms with higher powers of $x_0$ could also be included if they have any significance.
The detuning parameter is defined by $\delta_{1,0} = \omega_1 - \omega_0$.  We assume that the detuning is small compared to the fundamental frequency: $\delta_{1,0}/\omega_0 \ll 1$.  Both the $C_1$ and $C_2$ terms can generate a component that is a constant times $x_1$.  In this small detuning limit, the effect of this term is a simple shift in the detuning to the new value
\begin{equation}\label{deltashift}
\delta= \delta_{1,0}+ \langle x_0 \rangle C_1+ \langle x_0^2 \rangle C_2.
\end{equation}
Note that $\langle x_0 \rangle$ will typically be near zero when $x_0$ is approximately sinusoidal.  For convenience we rescale time so that the fundamental frequency $\omega_0 = 1$ and we define the net coupling coefficient $C$ as the combined amplitude of second harmonic components divided by $4$.  The resulting equation for the dynamics of $x_1$ is:
\begin{equation}\label{nonrad1}
\ddot{x}_1 = - [(1 + \delta)^2 + 4C  \cos(2 t)] x_1
\end{equation}
This equation has near-resonant terms that can lead to a slow evolution of the amplitude of $x_1$.  Usually such equations are reduced by an averaging procedure to equations that show the slow dynamics of the envelope of the oscillation and also its relative phase.  For example, this is the case in \citet{Buchler2011} where fast oscillating Equation set (1) is converted to the slow dynamics in Equation set (2).  However, in the current case, it is easier and more useful to solve the fast equation.  We try a solution of the form:
\begin{equation}\label{x_1}
x_1 = A_{1a} \cos[(1+\omega_B)t] - A_{1b} \cos[(1-\omega_B)t],
\end{equation}
where $\omega_B$ is the Blazhko frequency and $A_{1a}$ and $A_{1b}$ are the upper and lower side-peak amplitudes respectively.  As is shown below, the parameters of the problem will determine the ratio of the two amplitudes and also the Blazhko frequency.  While the fundamental frequency itself does not appear in the dynamics of $x_1$, it is of course present in $x_0$.  Note that $\omega_B$ is not necessarily equal to the detuning $|\delta|$.  This is because the nonlinear interaction can shift the average frequency of the nonradial mode, pulling both side-peaks in towards the fundamental.  Substituting the expression for $x_1$ into Equation~\ref{nonrad1} and using a trigonometric identity to expand the cosine product, one obtains the following equation relating the coefficients of the terms with frequency $1+\omega_B$:
\begin{equation}\label{rightpeak}
A_{1a}(1+\omega_B)^2 = A_{1a}(1+\delta)^2 - 2A_{1b}C
\end{equation}
and for $1-\omega_B$ one obtains:
\begin{equation}\label{leftpeak}
A_{1b}(1-\omega_B)^2 = A_{1b}(1+\delta)^2 - 2A_{1a}C.
\end{equation}
There are also terms of frequency $3+\omega_B$ and  $3-\omega_B$ generated by the cosine product, however because these terms are far from resonance they can be neglected provided the detuning $\delta$ is small.  Off resonance terms are always dropped from amplitude equations because they do not stay in phase long enough to have much impact on the slow averaged dynamics.  For example, in \citet{Buchler2011} Equation (1) for $db/dt$ there is a resonant coupling term proportional to $a^9b^*$, but there are dozens of lower order non-resonant coupling terms that are simply ignored and omitted from the equations (e.g. terms proportional to $a^2b^2$, $a^{*3}b^2$, $a^3b^{*2}b$, etc.)  It was verified numerically that the peaks at  $3+\omega_B$ and  $3-\omega_B$ are small and go to zero in the limit of small $\delta$.

By also neglecting as small the quadratic terms $\omega_B^2$ and $\delta^2$, one can solve for $\omega_B$ obtaining:
\begin{equation}\label{omega_B}
\omega_B = \sqrt{\delta^2-C^2}.
\end{equation}
We will adopt the convention that $\omega_B$ is always positive, even when the adjusted detuning $\delta$ is negative.  Note that the solution for $\omega_B$ only exists if $C<|\delta|$.  (Recall that $C$ is nonnegative.)  One can also solve for $R$, the ratio of the amplitudes of the side-peaks.  We will define $R$ differently depending on the sign of $\delta$.  For positive $\delta$ we define $R=A_{1b}/A_{1a}$ and for negative $\delta$ we define $R=-A_{1a}/A_{1b}$.  One finds with these definitions that $R$ is restricted to the range $0 \le R<1$ and is given by:
\begin{equation}\label{R}
R=|\delta/C|-\sqrt{|\delta/C|^2-1}.
\end{equation}
Note that Equation~\ref{nonrad1} is a second order linear homogeneous differential equation with variable coefficients.  Such equations are expected to have two linearly independent solutions.  The second solution is obtained by replacing the cosines in Equation~\ref{x_1} with sines and changing the sign of the second term to plus.  This leads to the same expressions for $R$ and $\omega_B$.  The most general solution is a linear combination of the sine and cosine forms.

\begin{figure}
\begin{center}
\includegraphics[width=1.0\columnwidth]{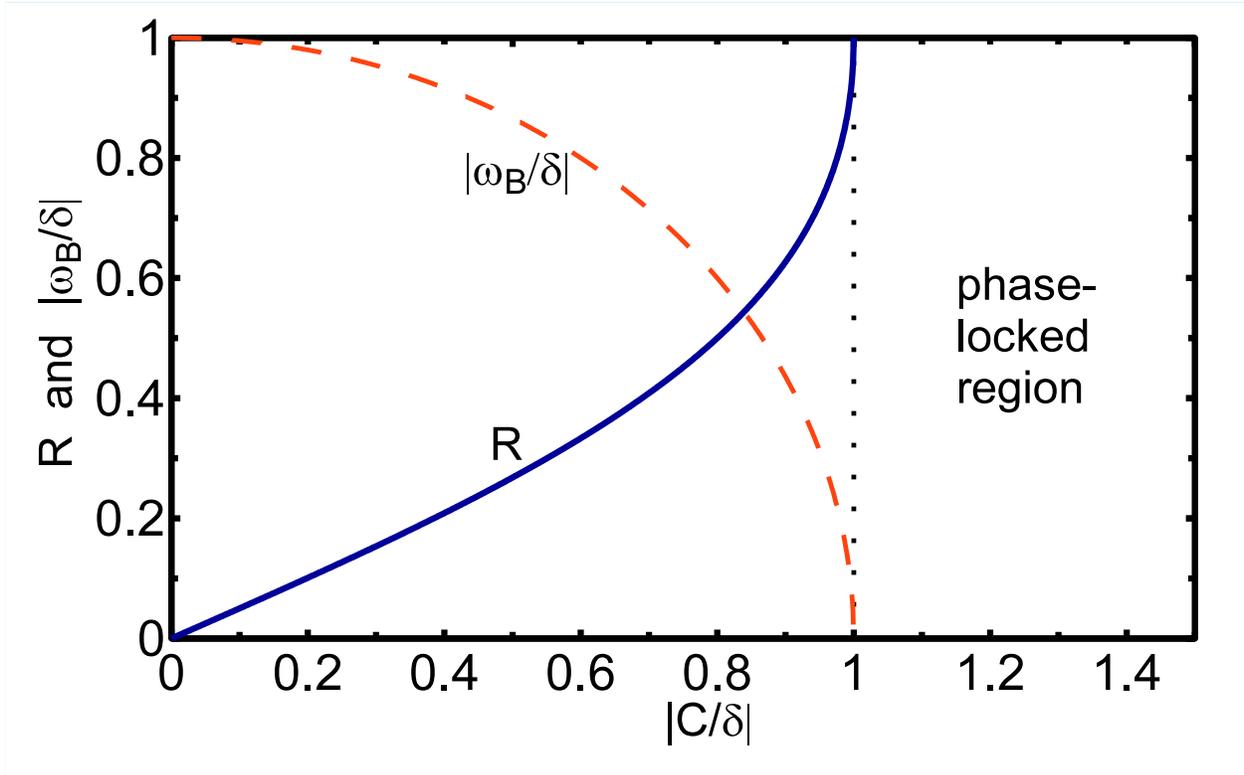}
\caption{\label{fig_dfix}Shows $R$ and $|\omega_B/\delta|$ as a functions of $|C/\delta|$, where $R$ is the side-peak amplitude ratio, $C$ is the coupling strength, $\delta$ is the adjusted detuning (see Equation~\ref{deltashift}) and $\omega_B$ is the Blazhko frequency.  For $|C/\delta| > 1$ the modes become phase-locked and there are no side-peaks (assuming the phase-locked state is stable).  $R=1$ is a perfectly symmetric triplet with both side-peaks of equal height.  As $R$ is reduced toward zero the triplet becomes increasingly asymmetric.}
\end{center}
\end{figure}

Two approaches are useful to examining these equations.  In the first we assume that the $\delta$ has a fixed value and see what happens as the coupling strength $C$ is varied.  This is shown in Figure~\ref{fig_dfix}.
Note that the adjusted detuning $\delta$ is typically shifted from the original detuning $\delta_{1,0}$ in proportion to coupling strength according to Equation~\ref{deltashift}.  Note that as $C$ is increased toward $|\delta|$, the Blazhko frequency decreases toward zero and the side-peak amplitude ratio approaches unity.  For $C > |\delta|$ phase locking occurs between the fundamental and the nonradial mode and a net rate of energy flow may occur between them.  At the other end of the range, when the coupling $C$ is decreased to zero, $\omega_B$ increases to become exactly equal to the detuning $|\delta|$ and the side-peak amplitude ratio $R$ goes to zero.  In the weak coupling limit, i.e. for $C \ll |\delta|$, Equation~\ref{R} simplifies to:
\begin{equation}\label{RsmallC}
R \approx |C/2\delta|
\end{equation}

\begin{figure}
\begin{center}
\includegraphics[width=1.0\columnwidth]{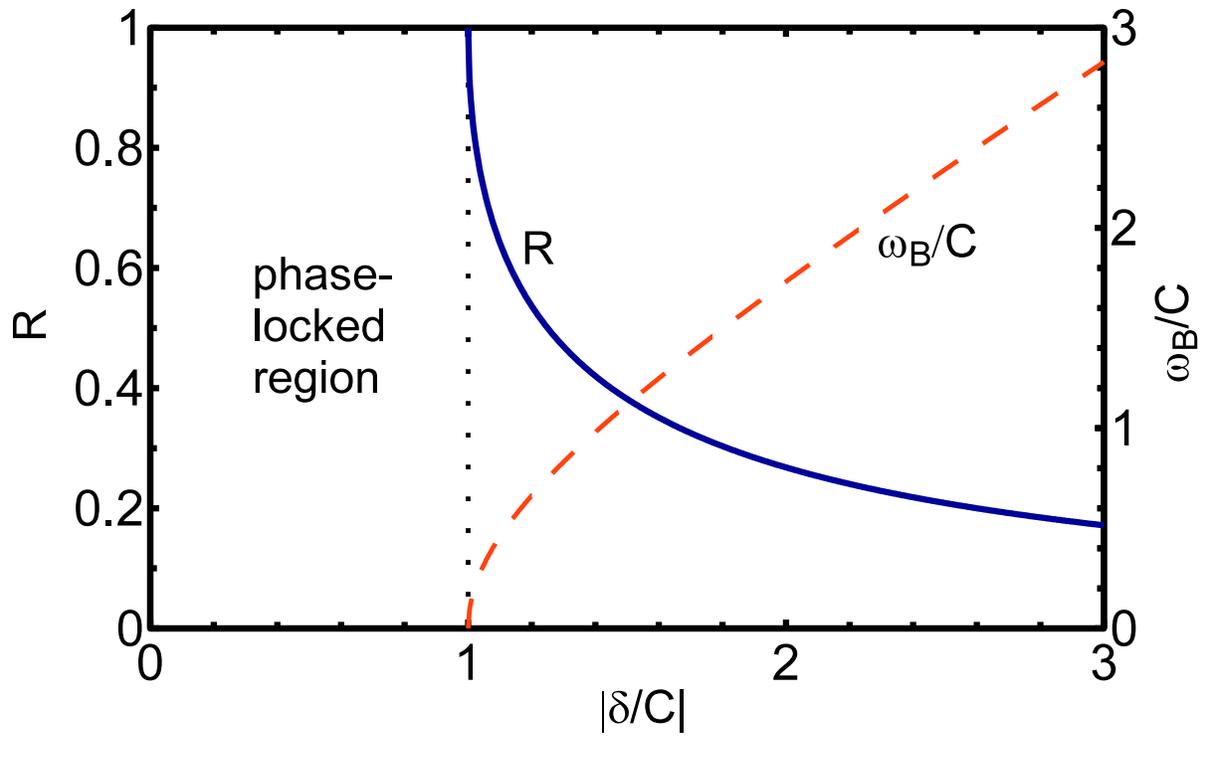}
\caption{\label{fig_cfix}Shows $R$ and $\omega_B/C$ as functions of $|\delta/C|$, where $R$ is the side-peak amplitude ratio, $C$ is the coupling strength, $\delta$ is the adjusted detuning  (see Equation~\ref{deltashift}) and $\omega_B$ is the Blazhko frequency.  For $|\delta/C| < 1$ there is phase-locking, which (if stable) will result in a spectrum containing the fundamental and its harmonics only, with no side-peaks.  For $|\delta|$ slightly greater than $C$, the Blazhko frequency emerges from zero and is considerably smaller than the detuning.  The triplet that appears in the spectrum is nearly balanced in side-peak amplitudes.  As the detuning is increased, the side-peak ratio decreases toward zero making the triplet increasingly asymmetric and the Blazhko frequency starts to approach the value of the detuning.}
\end{center}
\end{figure}

In the second approach we assume that the coupling strength is fixed and examine what happens if the detuning is changed.  This is shown in Figure~\ref{fig_cfix}.
Since the detuning can be negative, the phase-locked region actually extends from $\delta = -C$ to $+C$.  In terms of the unadjusted detuning $\delta_{1,0}$ this range is offset as determined by Equation~\ref{deltashift}, and it is worth noting that if the offset is greater than $C$ then the phase-locking range will not include $\delta_{1,0}=0$.  If there is a cluster of peaks in the vicinity of the fundamental, we would expect the interaction to cause the peaks with the smallest adjusted detuning to become phase-locked and ``disappear" from the spectrum, with the remaining peaks now generally closer together, each with its own side-peak on the opposite side of the fundamental.

\begin{figure}
\begin{center}
\includegraphics[width=1.0\columnwidth,height=6in,keepaspectratio]{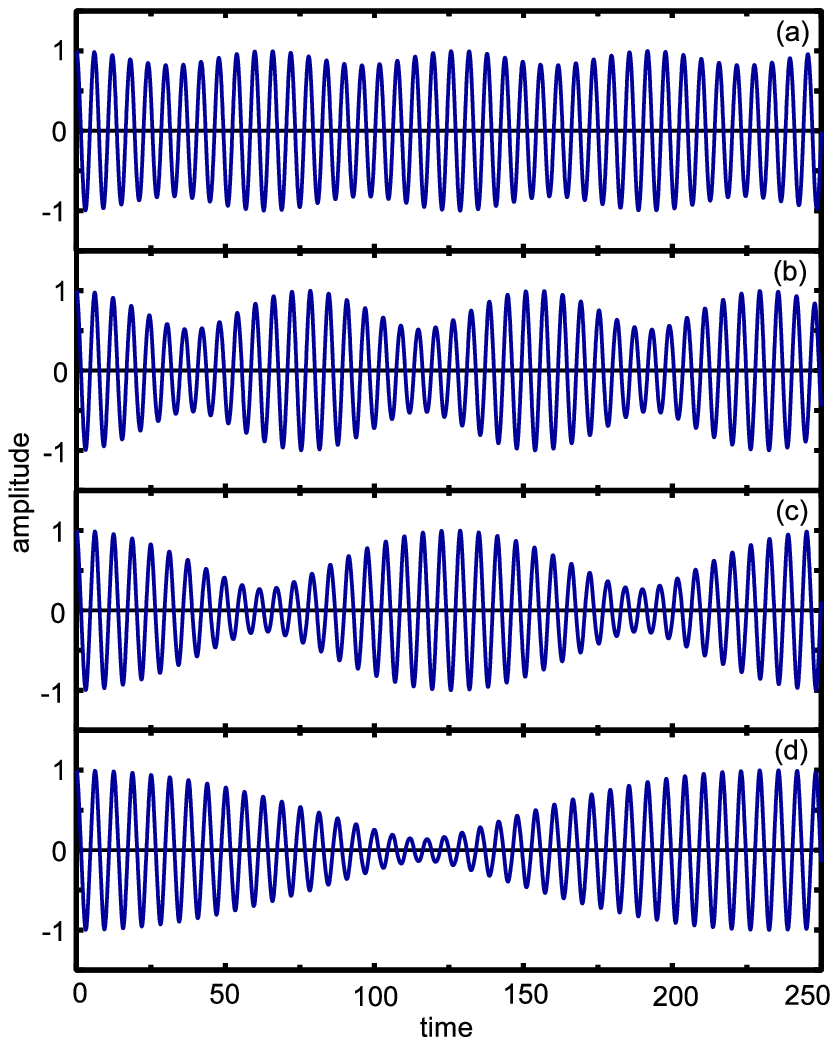}
\caption{\label{fig_adiawave}Shows $x_1$ as a function of time for different values of the coupling strength.  All have detuning $\delta=0.05$.  Curve (a) has $C=0.01$, (b) has $C=0.03$,  (c) has $C=0.045$ and (d) has $C=0.05$.  Phase-locking occurs for $C>0.05193155$.  The apparent modulation is caused by the interaction with the fundamental that is constantly changing due to the changing relative phase between the two modes.  Note the increasing depth of the modulation and the decreasing modulation frequency as the coupling is increased.}
\end{center}
\end{figure}

\begin{figure}
\begin{center}
\includegraphics[width=1.0\columnwidth]{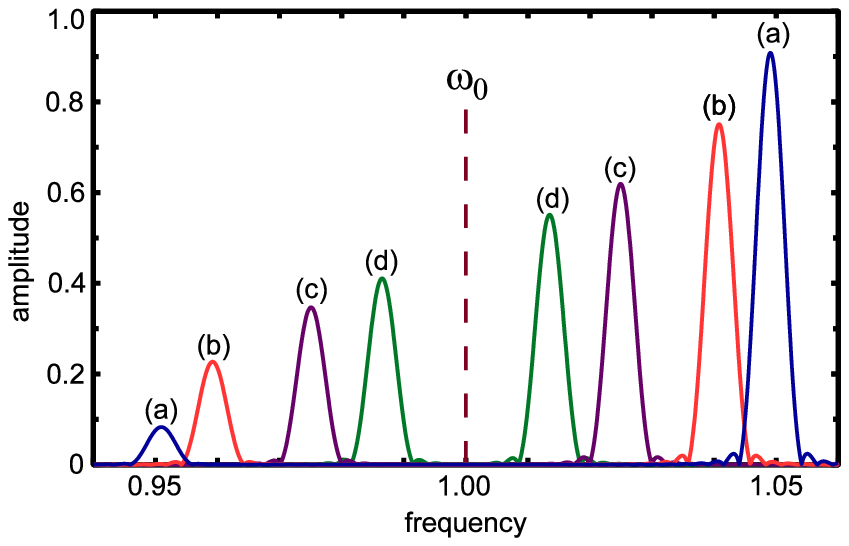}
\caption{\label{fig_adiaspec}Shows spectra for $x_1$ corresponding to the waveforms shown in Figure~\ref{fig_adiawave}.  The four cases are superimposed for comparison.  Each case consists of one left and one right side-peak of the fundamental labeled (a) through (d) in correspondence with Figure~\ref{fig_adiawave}.  Note that as the coupling strength is increased, the peaks move inwards and the amplitude ratio approaches 1.0.  Note that the fundamental is not present in the spectra of $x_1$.}
\end{center}
\end{figure}

We can also study Equation~\ref{nonrad1} numerically.  In Figure~\ref{fig_adiawave} we show $x_1(t)$ for a detuning $\delta=0.05$ and several values of the coupling strength $C$.
The phase-locking transition point was found numerically to be about $0.05193155$.  So it is about 4\% higher than the expected value of 0.05.  The discrepancy is due to the fact that $\delta$ has a finite but small size and thus the analytic results are only approximately correct.  Note that the waveform shown is only for the nonradial mode $x_1$ and does not include the fundamental $x_0$.  These will be combined together in some way in the real star to produce the observed light curve.  The modulated appearance of $x_1$ is due to its interaction with the fundamental.  The corresponding spectra are shown in Figure~\ref{fig_adiaspec}.
Only the side-peaks are present in the spectra of $x_1$; the fundamental is absent.  The modulation that is observed is at twice the Blazhko frequency because this is the spacing between the two side-peaks.

\subsection{Modes with Nonzero $m$}
\label{nonzerom}
Now we will briefly discuss the case of nonzero $m$.  Due to stellar rotation, these modes take the form of traveling waves that circulate around the star parallel to the equator.  By convention, negative values of $m$ correspond to prograde modes which travel in the same direction as the stellar rotation, while positive values of $m$ correspond to retrograde modes.  The rotation also causes a splitting of the frequencies as a function of $m$ which is reminiscent of the well known Zeeman splitting in atomic spectra:
\begin{equation}\label{split}
\omega_m = \omega_0 - m \lambda,
\end{equation}
where $\lambda$ is a mode dependent constant.  If one approximates the stellar rotation rate $\Omega$ as a constant (ignoring any dependence of $\Omega$ on $r$ and $\theta$) then it has been found \citep[see e.g.][Section 19.2 and Equation 19.45]{Unno} that for moderate to large values of the radial index $n$, p-modes have $\lambda \approx \Omega$ and g-modes have $\lambda \approx \{1-1/[l(l+1)]\}\Omega$.  So in general, $\lambda$ is expected to be in the range $\Omega/2$ to $\Omega$, with the low end of the range corresponding to dipolar g-modes.  One of these traveling wave modes by itself cannot couple resonantly to the fundamental.  However, a pair with opposite values of $m$ can couple to the fundamental and if phase-locked to the fundamental can extract energy from it.  It will generate a triplet in the spectrum and as such is a potential explanation for the Blazhko effect \citep{Nowakowski2001, Bryant2015}.  Due to the frequency splitting, the sum of the pair produces a standing wave pattern that slowly rotates at the frequency $\lambda$.  If not phase-locked to the fundamental and if viewed in a frame of reference that co-rotates at frequency $\lambda$, this combined mode can interact with the fundamental in the same manner described in this section.  In the nonrotating frame however, the observed light curve will appear amplitude modulated at the frequency $m \lambda$ as a result of the changing portion of the stellar surface that is visible from the direction of observation.  If phase-locked to the fundamental this will create a triplet spectrum, otherwise this will create a more complex spectrum.

It is also worth noting that it is possible, at least in theory, for a pair with opposite values of $m$ to produce a triplet spectrum with strong amplitude asymmetry.  This was demonstrated in \citet{Buchler1995} using amplitude equations.  Since saturation effects limit the energy that the pair can acquire through the opacity mechanism, there are nonlinear environmental-type terms appearing in the amplitude equations that generate this limiting effect.  Apparently the cross terms between the modes can result in one mode being strongly favored over the other even when the various coefficients for the two modes are only slightly different, i.e. the various coefficients in their Equation set 3.  However, energy can also be acquired by the pair through nonlinear resonant-type coupling to the fundamental mode (assuming phase-locking), and this process favors the pair of modes having equal amplitudes.  To the extent that this process dominates, the asymmetry between the pair will be diminished toward zero.

\subsection{Effect of the Interaction on the Fundamental}
\label{x0effects}
Retaining the approximation that the fundamental is very robust compared to the nonradial mode, we can ask what is the effect of the weak coupling terms from $x_1$ on the waveform of $x_0$.  Since we know from the observed the light curve and radial velocity waveforms for RR Lyrae stars that $x_0$ is operating in a highly nonlinear regime, we will discuss the effects qualitatively.  The two coupling terms coming from the potential that would appear in the equation for $\ddot{x}_0$ are $-C_1 x_1^2$ and $-2C_2 x_0 x_1^2$.  These terms are assumed to be very weak compared to their counterparts in the equation for $\ddot{x}_1$ because of our assumption of a robust fundamental which leads to $\langle x_0^2 \rangle$ being substantially larger than $\langle x_1^2 \rangle$.  The square of $x_1$ appearing in both of these includes the frequency $2\omega_0 + 2\omega_B$ and, when the opposing side-peak is present in $x_1$, also the frequency  $2\omega_0 - 2\omega_B$.  So the $C_1$ term may put small peaks at those two frequencies in the spectrum of $x_0$.  The $C_2$ term contains an additional factor of $x_0$ and thus it may generate small peaks in the $x_0$ spectrum with frequencies  $\omega_0 + 2\omega_B$ and  $3\omega_0 + 2\omega_B$ and, when the opposite side-peak is present in $x_1$, also the frequencies $\omega_0 - 2\omega_B$ and  $3\omega_0 - 2\omega_B$.  Since both $x_0$ and $x_1$ have an effect on the observed light curve, these peaks may be visible in the spectrum if they are large enough to appear above the noise.

One can also construct a dynamical model that includes both modes explicitly, i.e. one that has equations for both $\ddot{x}_0$ and $\ddot{x}_1$.  This model, with damping effects also included, is presented and discussed in Section~\ref{x0x1}.

\section{DISSIPATIVE RESULTS}
\label{dissipative}
Real stars have excitation and damping, and the equations of motion then contain forcing terms dependent on the velocity.  The linear term in the velocity can result in excitation or damping depending on the sign of the coefficient.  If there is excitation then some nonlinear damping may be required to prevent the mode from increasing without bound.  This could for example take the form of a cubic term in the velocity of the mode in question.  Another possibility (not applicable to the current problem) is that the excess energy may be transferred to a resonant mode where it is dissipated.

For the current problem, analytic analysis becomes rather difficult or impossible with the additional terms in the equations, so we will explore the results numerically.  The dynamics are started from an initial condition and then allowed to settle onto the attractor before the results are analyzed.  To Equation~\ref{nonrad1} we add a linear excitation term and a cubic damping term:
\begin{equation}\label{nonrad2}
\ddot{x}_1 = - [(1 + \delta)^2 + 4C  \cos(2 t)] x_1 +2\kappa \dot{x}_1 +Q \dot{x}_1^3
\end{equation}
Where $\kappa$ is the growth rate coefficient.  It must be positive to make the mode unstable.  $Q$ is the cubic damping coefficient.  It must be negative to restrain the system from unlimited growth.  As it is the only nonlinear term in the equation, it is possible to rescale the mode amplitude in a way that makes $Q = -1$.  As discussed in Appendix~\ref{complex} and also verified numerically, there is only one other possible quadratic or cubic term that can act to restrain the growth of the linear excitation term.  This term would have the form $Q|x_1|^2\dot{x_1}$ and it will generate nearly identical dynamics when $\delta$ is small (though with a different scale factor).

\begin{figure}
\begin{center}
\includegraphics[width=1.0\columnwidth]{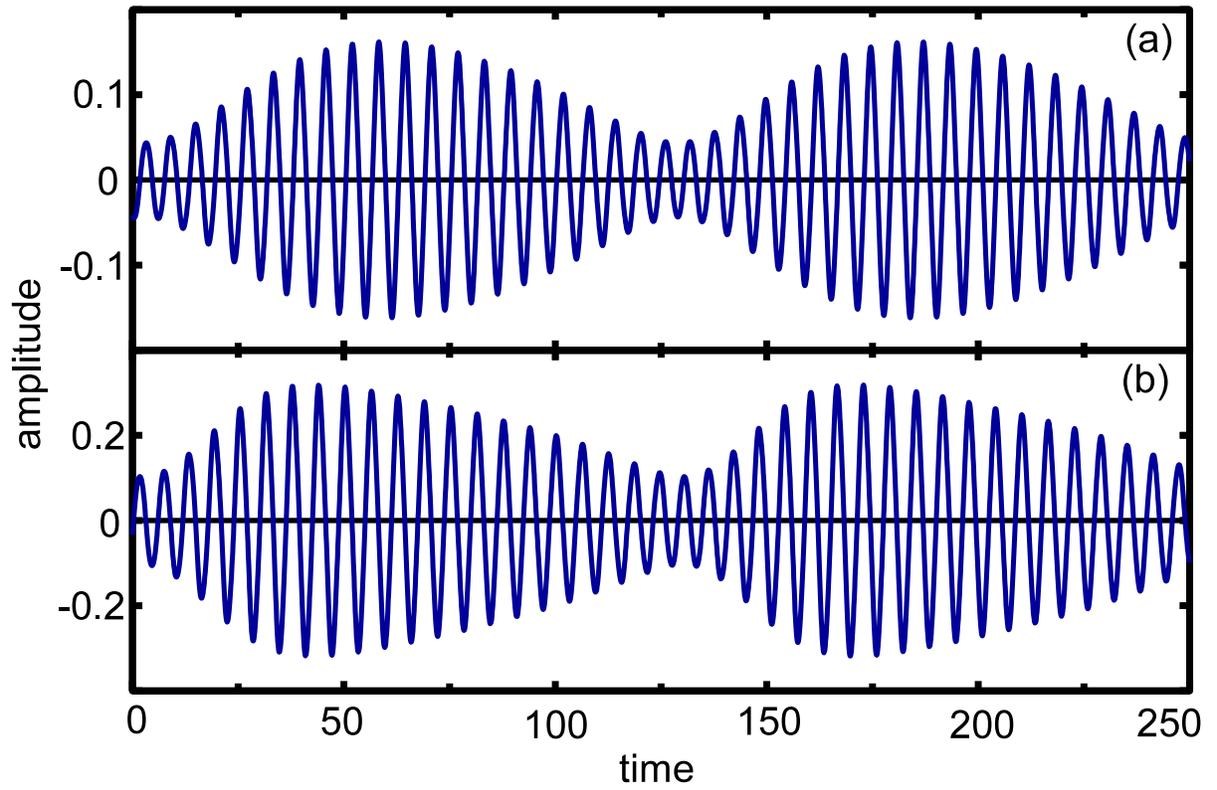}
\caption{\label{fig_disswave}Shows waveforms for two values of the growth rate $\kappa$.  Note the change in the modulation envelope.  Both cases have detuning $\delta=0.05$, coupling $C=0.045$ and nonlinear damping coefficient $Q=-1$.  (a) $\kappa=0.005$  (b) $\kappa=0.02$.  Compare these to the results for $\kappa=0$ given in Figure~\ref{fig_adiawave}(c).}
\end{center}
\end{figure}

\begin{figure}
\begin{center}
\includegraphics[width=1.0\columnwidth,height=6in,keepaspectratio]{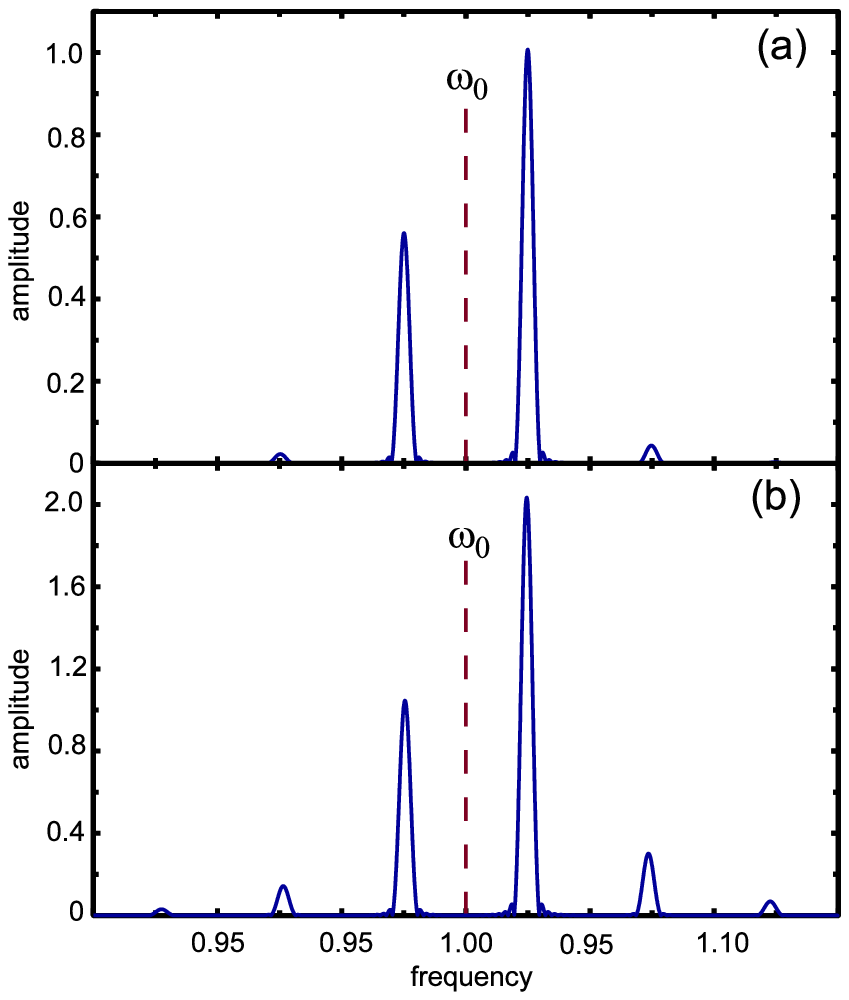}
\caption{\label{fig_dissspec}Shows spectra corresponding to the waveforms shown in Figure~\ref{fig_disswave} for two values of the growth rate $\kappa$.  Note the emergence of the third and fifth order side-peaks as $\kappa$ is increased.}
\end{center}
\end{figure}

In Figure~\ref{fig_disswave} we show time series for Equation~\ref{nonrad2} with two values of the growth rate $\kappa$.
The detuning $\delta$ and coupling $C$ fixed at the same values as used previously in Figures~\ref{fig_adiawave}(c) and~\ref{fig_adiaspec}(c).  As can be seen, in the small $\kappa$ limit the dynamics approach the adiabatic results.  As $\kappa$ is increased, we see the emergence of additional side-peaks in the spectrum and a distortion of the modulation envelope giving it somewhat of a saw-tooth character.  The spectra for the same values of $\kappa$ are shown in Figure~\ref{fig_dissspec}.
The nonlinearity in the equations generates nonlinear mixing products between the first order side-peaks.  As a result, the additional side-peaks so created are also spaced by twice the Blazhko frequency, i.e. this will only generate side-peaks of odd order.  Evidence for this can be found in observational data; see Section \ref{odd} for discussion of results from V808 Cyg.

Most previous Blazhko models involve some process that results in the actual modulation of the fundamental mode at the Blazhko frequency and it is this modulation that produces the observed effect.  There appears to be no obvious reason for this modulation to favor odd-order side-peaks.

\subsection{Explicit Modeling of the Fundamental}
\label{x0x1}
We can explicitly include the fundamental by adding an equation for $\ddot{x}_0$.  Although this mode is strongly non-sinusoidal near the surface, one approach is to nevertheless use a simple model that produces a sine wave that may slowly change in response to the interaction.  This is what is usually done, \citep[see e.g.][Equation Set~1]{Buchler2011}.  One might try to justify this by claiming that most of the energy of the fundamental is far below the stellar surface, and that the oscillation is relatively sinusoidal in that region.  For our current case the resulting equations are:
\begin{equation}\label{rad1b}
\ddot{x}_0 = - [1 + 2C_2 x_1^2] x_0 +2\kappa_0 \dot{x}_0 +Q_0 \dot{x}_0^3
\end{equation}
\begin{equation}\label{nonrad1b}
\ddot{x}_1 = - [(1 + \delta_{1,0})^2 + 2C_2  x_0^2] x_1 +2\kappa_1 \dot{x}_1 +Q_1 \dot{x}_1^3
\end{equation}
As before, time has been scaled so that the frequency of the fundamental mode (mode~0) is exactly 1.0 and $ \delta_{1,0}$ is the unadjusted detuning of the nonradial mode (mode~1).  The $C_1$ term is not used because it is incompatible with our modeling scheme for the fundamental, i.e. lacking strong second harmonic content in the fundamental, this term is nonresonant and will have no impact.  Deriving the $C_2$ terms from a potential guarantees that energy leaving the fundamental via this term is transferred to the nonradial mode and vice versa.  To be near the limit of a robust fundamental we simply need to choose parameters such that $\langle A_0 \rangle \gg \langle A_1 \rangle$, where $A_0$ and $A_1$ are the amplitudes of the two modes which may be slowly varying because of the interaction between them.  Provided $\kappa_0$ is small, mode~0 should oscillate with an approximate sine wave of amplitude $A_0$, and $x_0^2$ which appears in Equation~\ref{nonrad1b} will have a second harmonic component of amplitude $A_0^2/2$.  Thus in the robust fundamental limit, the dynamics of $x_1$ should approximately match the dynamics of the earlier results provided we make the connection: 
\begin{equation}\label{C_2vsC}
C_2=4C/A_0^2
\end{equation}
So the range of values of $\delta$ that results in phase-locking is from $-C_2A_0^2/4$ to $+C_2A_0^2/4$.  With $x_0$ being approximately a sine wave, we have $\langle x_0^2 \rangle = A_0^2/2$, and therefore from Equation~\ref{deltashift} we have $\delta_{1,0} = \delta - C_2A_0^2/2$.  So the range of values of $\delta_{1,0}$ that results in phase-locking is from $-3C_2A_0^2/4$ to $-C_2A_0^2/4$ which does not include zero.  The lowest Blazhko frequencies should occur just outside this range.

Based on the motion being an approximate sine wave, one can solve explicitly for the amplitude by calculating the energy input each cycle by the growth rate and setting it equal to the energy loss through nonlinear damping.  This yields:
\begin{equation}\label{amplitude}
A^2= -8\kappa/(3Q\omega^2)
\end{equation}
This equation could be applied to $A_1$ as well as $A_0$, but in the case of significant coupling, $A_1$ will be varying with time and so the equation may then provide a rough estimate of the average value of $A_1^2$.

As a test of the previous results the following parameter values were chosen:  $\kappa_0=0.005$, $\kappa_1=0.005$, $C_2=0.045$ and $Q_1=-1.0$.  For convenience $Q_0$ was set to $(2/3)\kappa_0 = 0.00333333$, which by Equation~\ref{amplitude} sets $A_0\approx 2.0$ and by Equation~\ref{C_2vsC} sets $C \approx C_2 = 0.045$.  Numerically integrating the equations, with $\delta_{1,0}$ set to $-0.04$, we obtain a close match to the dynamics shown previously in Figure~\ref{fig_disswave}(a).  By Equation~\ref{deltashift} this corresponds to an adjusted detuning $\delta=0.05$ so the parameter values are all matched to that previous case, thereby verifying that the approximations used in the simplified equations were valid.  The effect of the interaction on the fundamental is very slight as expected, the modulation envelope of $x_0$ having an amplitude of about 0.003 or about 0.15\% of $\langle A_0 \rangle$.  For a slightly more negative $\delta_{1,0}$, mode~1 becomes phase-locked to mode~0, extracts energy from it and increases in amplitude.  So, for example, for $\delta_{1,0} = -0.1$, mode~1 is phase-locked and has a constant amplitude of $A_1=0.30$, whereas the maximum amplitude achieved for $\delta_{1,0} = -0.04$ is about 0.163.  The fundamental is also loaded down slightly now having an amplitude of about 1.83, down from its original value of 2.00.  Since this a stable locking of the two modes, there is no modulation and a spectrum would show only a single peak at the fundamental frequency.  Thus it would be difficult to determine from observational data that mode~1 was excited.  Going still more negative in the detuning, phase-locking is again lost but this time with the adjusted detuning being negative.  One again sees dynamics closely matching Figure~\ref{fig_disswave}(a) with $\delta_{1,0} = -0.153$.

More can be done with these equations, in particular, one can see how the behavior changes as the amplitude of the nonradial mode is allowed to be closer to that of the fundamental.  From a preliminary examination it appears that qualitatively similar results may persist even in cases where $A_1$ is on the order of half the value of $A_0$.  This may warrant more study at a future date.

\section{OBSERVATIONAL RESULTS FROM V808 CYG}
Two interesting pieces of evidence in support of the model were found in the spectra of V808 Cyg: a preference for odd order side-peaks of the fundamental and a mode that appears to be modulated at twice the Blazhko frequency.  These are discussed below in Sections~\ref{odd} and~\ref{modulation} respectively.  Results given below were calculated from \emph{Kepler} project KIC 4484128, Q1 through Q17, long cadence corrected flux data using Period04 software.  The fundamental frequency $f_0$ is about 1.825273 cycles per day.  The Blazhko frequency $\omega_B$ is about 0.01085.  Average (or zero point) flux for this data is about 8835.  Flux of the fundamental is 2857 and the first order side-peaks, left and right, are 539 and 463 respectively.  This is about 1460 days or 4 years worth of data.  The time-step is 0.020435 days and the Nyquist frequency is 24.4708 which falls between the 13th and 14th harmonics of the fundamental.  The full width at half maximum (FWHM) for the spectral peaks is about 0.0008.
\subsection{Odd Side-peaks Preferred}
\label{odd}
\begin{figure}
\begin{center}
\includegraphics[width=1.0\columnwidth,height=5in,keepaspectratio]{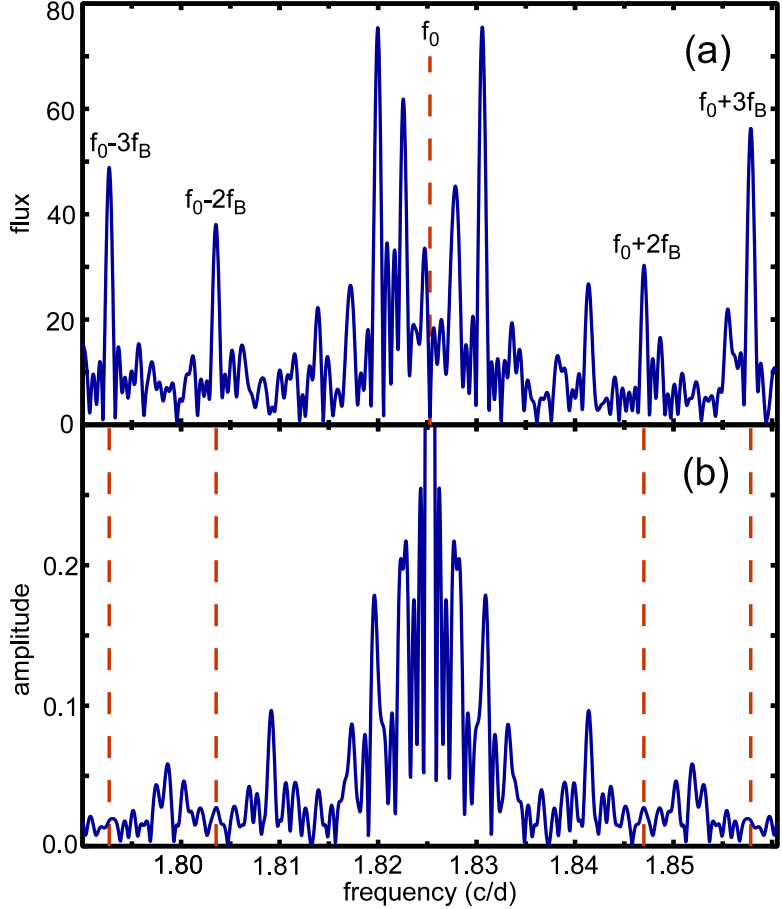}
\caption{\label{fig_f0sidepeaks}Part (a) shows septuplet spectrum for V808 Cyg.  The inner triplet spectral peaks including the fundamental itself have been removed leaving only the second and third order side-peaks on each side as indicated by the labels.  Note that the second order peaks are smaller than the third order peaks when they might be expected to be much larger.  This is consistent with the effects of nonlinear damping on $x_1$ which generates odd-order side-peaks (see text and Figure~\ref{fig_dissspec}).  The large central peaks (unlabeled) are windowing artifacts, i.e. they result from the modulation of the fundamental by the ``window" made up of the discontinuous chunks of data available for analysis.  Part (b) shows the window spectrum.  It is offset in frequency by $f_0$ in order to show the effect of the window modulation on that peak.  The artifacts would be larger if the fundamental had not been removed.  The remnant is presumably due to slight variations in the frequency and amplitude of $f_0$ over the time of the data set.}
\end{center}
\end{figure}

V808 Cyg has relatively little asymmetry which indicates according to the theory that there may be a fairly strong interaction between Mode 1 and mode 0 (the fundamental).  As discussed in Section \ref{dissipative}, stars with a strong interaction between the modes may tend to favor the generation of odd order side-peaks.  This is in fact found to be the case for V808 Cyg as shown in Figure~\ref{fig_f0sidepeaks}, which strongly favors the third order side-peaks over the second order.
The third order should be much smaller but are instead somewhat larger than the second order.  It should be noted that there is at least two other mechanisms that can generate these peaks including the second order ones.  The first is the nonlinear light-curve function (or functional) that transforms the dynamical variables of the excited modes into the observed light curve.  This function (or functional) is unknown but is observed in some cases to be approximated by a nonlinear function of the radial velocity \citep[see e.g.][]{Bryant2015}.  This can generate nonlinear mixing products between mode~0 (the fundamental) and mode~1 including side-peaks of any order.  The second mechanism comes from the effect of the interaction on the fundamental as discussed in Section~\ref{x0effects}.  In the adiabatic case this will put small second order side-peaks in the spectrum.  In the dissipative case it can also put higher even-order peaks in the spectrum.  The more robust the fundamental is compared to the nonradial mode, the smaller these even-order side-peaks will be.  Neither of these other mechanisms can explain the observed preference for the odd-order side-peaks.

The fourth and fifth order peaks (not shown) are only marginally detected, being less than a factor of two higher than neighboring noise bumps.  The amplitude of the fourth order are down by about a factor of 5 from the third order, while the fifth order are down about another 30\%.  This again indicates a preference for the odd-order cases since one might otherwise expect another factor of 5 drop in amplitude between the fourth and fifth order.  The other mechanisms mentioned above may be playing a more significant role as the order increases.  Higher order peaks are not visible.  The measured peak amplitudes are given in Appendix~\ref{freqtab}.

\subsection{Modulation of Additional Modes}
\label{modulation}
\begin{figure}
\begin{center}
\includegraphics[width=6.0in]{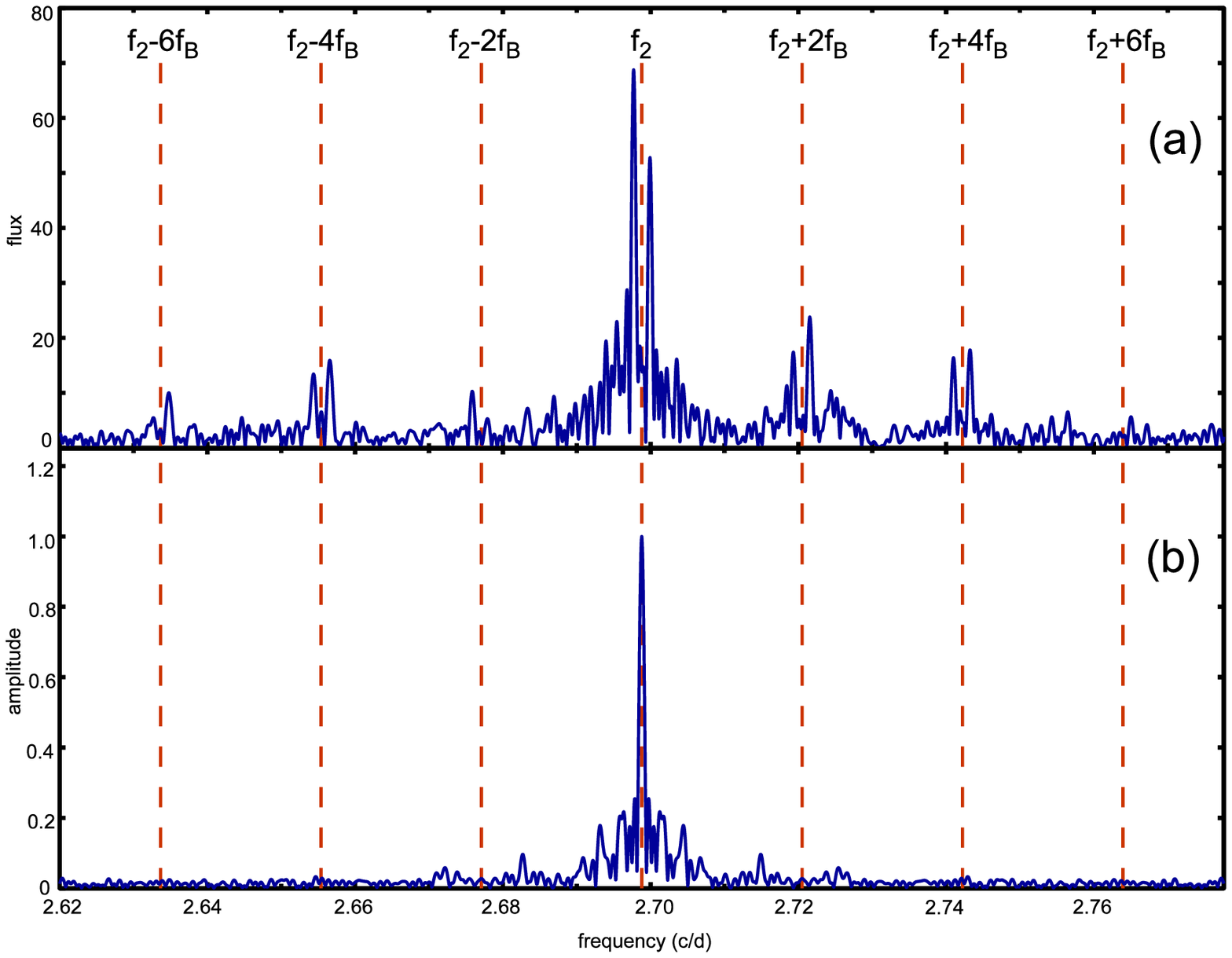}
\caption{\label{fig_mod}Part (a) shows $f_2$ a modulated peak found in the spectrum for V808 Cyg.  Note that the side-peaks are displaced from the main peak by multiples of twice the Blazhko frequency.  This double Blazhko modulation is shown to be consistent with the interactive-beating-modes model of the Blazhko effect (see text).  Note that the peaks are split, possibly indicating two closely spaced modes or a mode split by stellar rotation and also indicating that at least one and probably both peaks correspond to nonradial modes.  Part (b) shows the window spectrum.  Its frequency is offset by $f_2$ in order to show that window modulation is the source of the noise skirt at the base of this peak in part (a).}
\end{center}
\end{figure}

An interesting peak is to be found in the spectrum of V808 Cyg.  This peak appears to have modulation side-peaks as shown in Figure~\ref{fig_mod}.
It is close in frequency to $(3/2)f_0 = 2.7379$ but is displaced from it by about 50 times the FWHM of the spectral peaks.  We therefore assume that the peak is generated by an excited mode which we will refer to as mode~2 with frequency $f_2$ and amplitude $x_2$.  We will continue to refer to the fundamental as mode~0 and the mode that is near-resonant with it as mode~1.  Note that $f_2$ is actually split into two peaks with frequencies of $2.697753$ and $2.699816$.  One explanation for this split would be that the noisy turbulent environment of the star is causing the amplitude to fluctuate in a way that happens to look like the sum of these two frequencies.  A second explanation is that there really are two closely spaced modes that generate these peaks.  In favor of the second explanation is the fact that the difference between the peak frequencies is 0.002062 and from this one can determine that the two modes slipped in phase by about 3.0 cycles over the time span of the full data set (approximately 4 years of data from the Kepler project).  This seems like an excessive number of consistent cycles to be generated by a random process.  Also, when the data is divided into two parts of approximately 2 years each, the same split peaks are seen in both spectra with approximately the same amplitudes.  If the split peak is the result of random noise, then it would be quite unlikely to get the same spectra for two non-overlapping time intervals.  Thus there probably are two distinct modes involved here and since they are so close together it is reasonable to assume that at least one is a nonradial mode.

A third possibility, which is really a special case of the second possibility, is that the peak is split by stellar rotation, implying mode 2 is a nonradial mode with a nonzero index $m$.  Since the magnitude of the splitting is related to the rotation rate as discussed in Section~\ref{nonzerom}, this would imply, for $m=\pm 1$ a stellar rotation period for V808 Cyg of between 455 and 909 days depending on the characteristics of mode~2.  There is a constraint on rotation periods based on studies that analyze atomic line-widths.  These studies specify a maximum value for $v \sin i$, where $v$ is the rotational velocity at the equator and $i$ is the angle between the line of sight and the stellar rotation axis.  \citet{Peterson1996} and \citet{Chadid2013} give a value of 10km/s while \citet{Preston2013} give a value of 6km/s.  \citet{Smith1995}, Table 1.1, states that the radii of RR Lyrae stars are in the range of 4 to 6 solar radii.  If we choose $v$ = 10km/s and $r$ = 4 solar radii, we get a minimum acceptable value for the period of about 20 days.  This is well below the values indicated above for rotational splitting.

The side-peaks are also split into pairs of the same spacing.  A total of five pairs of side-peaks are visible, three on the left and two on the right.  The fact that it appears modulated is a strong indication that this peak (or pair of peaks) does correspond to an actual vibrational mode (or pair of modes) as opposed to the explanation that it is some kind of noisy period-doubling byproduct.  Note that any modulation of the fundamental in that case would necessarily produce side-peaks centered on $(3/2)f_0$ rather than centered on the offset frequency $f_2$ as is observed.  Reduced amplitude copies of the modulation pattern can be found at locations that are offset by multiples of $f_0$.  This shows that these frequencies are mixing with the fundamental and that therefore we can rule out the possibility of $f_2$ being from some other star in the background that appears in very close proximity to V808 Cyg.  Frequency tables for V808 Cyg are given in Appendix~\ref{freqtab}.

Perhaps even more interesting is that the apparent modulation frequency, based on the side-peak spacing, is twice the Blazhko frequency.  Based on measurements from the entire data set, the Blazhko frequency is about 0.01085, so twice this is 0.02170.  The side-peak spacing is determined to be about 0.02165 so this is a very accurate match.  We have an explanation for this doubled frequency in terms of the interactive-beating-modes model, i.e. the Blazhko model presented in this paper.  As explained below, the modulation of Mode~2 observed in the current problem may be due to the type of nonlinear interaction that is not phase-dependent.

The Blazhko effect generates an oscillatory environment that can have a modulatory effect on the frequency of Mode 2.  This can be seen through the ``$T$" terms that often appear in the complex amplitude equations, see, e.g. Equation (1) in \citet{Buchler2011}, Equations (1a) and (1b) in \citet{Moskalik1990} and Equations (70) and (71) in \citet{Buchler1984}.  (See Appendix~\ref{complex} of this paper for discussion of the complex equations).  In the equation for $da/dt$ in the cited papers, this term has the form $T|b|^2a$ where $T$ is a complex coefficient and $a$ and $b$ are the complex amplitudes of Modes A and B.  The imaginary component of $T$ results in the frequency of Mode A being shifted in proportion to the square of the amplitude of Mode B, while the real component of $T$ results in a similar shift in the growth rate of Mode A.  Note that this term does not depend on the relative phase of Mode A and will not cause phase-locking between the modes.  This is in contrast, for example, to the ``C" terms appearing in Equation (1) in \citet{Buchler2011} which can result in phase-locking between modes A and B.    So the presence of the $T$ term in the equation for $da/dt$ can be thought of as an effect of the ``environment" generated by Mode B on the dynamics of Mode A.  If $|b|$ is constant in time then the result is simply a constant shift in the oscillatory properties  of Mode A.  But if it is slowly varying the result will be to modulate Mode A.  Note that environmental terms are sometimes referred to as saturation terms, see e.g. Equations~6 and~7 in \citet{Nowakowski2001}.

Extending this analysis to the current problem we consider 3 cases.  In the first case we assume there is an interaction between modes 0 and 1 as discussed earlier in this paper which results in $x_1$ having a modulated amplitude as was shown in Figures~\ref{fig_adiawave} and~\ref{fig_disswave}, with corresponding spectra shown in Figures~\ref{fig_adiaspec} and~\ref{fig_dissspec}.  Since the fundamental does not appear in the spectrum of $x_1$, the modulation frequency is twice the Blazhko frequency (i.e. it is the spacing between the left and right side-peaks).  Thus a term of the form $T|z_1|^2z_2$ appearing in the equation for $\dot{z}_2$ will generate a shift in the frequency and/or growth rate that is varying at twice the Blazhko frequency.

A second case to consider is the combined effect of modes 0 and 1 on mode 2.  The lowest order terms of interest which might appear the equation for $\dot{z}_2$ are of the form $(U_1z_0^*z_1+U_2z_1^*z_0)z_2$, where $U_1$ and $U_2$ are complex coefficients.  The dominant frequencies of the two terms in the parentheses are $\omega_B$ and $-\omega_B$ respectively so these terms are slowly varying environmental terms that will modulate mode 2 at the Blazhko frequency.  However, there are symmetry conditions that apply in a lot of cases that will cause $U_1$ and $U_2$ to be exactly zero.  This occurs when mode 1 is antisymmetric under certain transformations.  Modes with $m=0$ and odd $l$ are antisymmetric under pole reversal, i.e. looking at the inverted star.  Modes with $m \ne 0$ are antisymmetric under rotation about the pole by $\pi/m$.  The environment for mode 2 is generated by the combined oscillations of modes 0 and 1 and as a result is slowly varying as the relative phase between modes 0 and 1 changes due to the Blazhko effect.  Due to the antisymmetry, if we move forward in time by one half Blazhko cycle the relative phase in the transformed coordinates will exactly match that of the original time with the original coordinates.  But this choice of coordinates cannot affect the dynamics and therefore the environmental effect must repeat every half Blazhko cycle and have period $2\omega_B$.  Thus $U_1$ and $U_2$ must be exactly zero for these cases.  The only cases which do not have such a symmetry are those with $m=0$ and even $l$.  Only these choices for mode 1 can result in modulation of mode 2 at once, rather than twice, the Blazhko frequency.  Note that as the degree $l$ increases, this distinction between odd and even diminishes in importance and so the $U$ coefficients for even $l$ should tend toward zero to match the odd results as $l$ is increased.  For the dipolar modes ($l=1$) and other modes with the antisymmetry condition we must use the next higher order term, $(U_1' z_0^{*2}z_1^2+U_2' z_1^{*2}z_0^2)z_2$, which will modulate at the frequency of $2\omega_B$.

A third case for modulation of mode 2 occurs if the Blazhko effect results from the split-pair model of \citet{Nowakowski2001} in which the Blazhko effect is produced by a phase-locking between the fundamental and a mode pair, designated 1a and 1b, with opposite values of $m \ne 0$ that is split into two slightly different frequencies by stellar rotation (see Section~\ref{nonzerom}).  These two modes form the two side-peaks of the fundamental.  For V808 Cyg, this would imply, for $m=\pm 1$, a stellar rotation period between 46 and 92 days depending on the characteristics of this mode.  Note that this is above the minimum of 20 implied by atomic line-width analysis (see discussion earlier in this Section).  This case can also produce the double Blazhko modulation, with terms in the equation for $\dot{z}_2$ of the form $(W_1z_{1a}z_{1b}^* +W_2z_{1a}^*z_{1b})z_2$.  Because of the symmetry properties mentioned above, the terms which could produce modulation at once rather than twice the Blazhko frequency are eliminated.

Other recent Blazhko models would appear to be incompatible with the observed double Blazhko modulation of mode 2.  These generally assume that the Blazhko effect involves some process modulating the fundamental mode at the Blazhko frequency.  The dominant environmental term in the equation for $\dot{z}_2$ would then be of the form $T|z_0|^2z_2$.  But since fundamental amplitude $|z_0|$ is fluctuating at the Blazhko frequency this will simply modulate mode 2 at that same frequency and not double it.

In the purely adiabatic limit, the terms discussed above can be derived from corresponding terms in the potential energy.  If, for example, we consider a term in the potential energy of the form $Ux_0x _1x_2^2$, where $U$ is a real coefficient, the equation for $\dot{x}_2$ will have a corresponding term $-2Ux_0x _1x_2$ (i.e. negative of the derivative with respect $x_2$).  Following the discussion in Appendix~\ref{complex}, the corresponding resonant terms in the equation for $\dot{z}_2$ are $(2iU/\omega_2)(z_0z_1^*+z_0^*z_1)z_2$.  Thus in this case, $U_1$ and $U_2$ as defined previously are equal and purely imaginary.  Since the expression multiplying $z_2$ is purely imaginary this corresponds to a pure frequency modulation.  Similar results apply to the $T$, $U'$ and $W$ coefficients; they are all purely imaginary with $U_1'=U_2'$ and $W_1=W_2$ and all result in frequency modulation.  However, if we are not close to the adiabatic limit, then all the coefficients should be regarded as independent complex numbers.

An interesting question is whether the symmetry properties of the dynamics are retained when the amplitude is large and therefore strongly nonlinear.  This may be indicated in the current case by the highly non-sinusoidal nature of the fundamental oscillation.  Studies of simpler nonlinear systems have indicated that symmetrical quasiperiodic oscillations may be quite common; see e.g. Figure (2b) in \citet{Bryant1987}.  As discussed in \citet{Bryant1984, Bryant1987} another possibility is complementary asymmetric attractor pairs, though this would seem unlikely except when phase locking occurs; see e.g. Figure (3b) in \citet{Bryant1987}.
 
\section{CONCLUSIONS}
\label{conclusions}
In this paper we have explored a new model of the Blazhko effect called the ``interactive beating-modes model".  The model is generated by including a nonlinear interaction between the two modes that make up a model described in a previous work \citep{Bryant2015}.  In that model the Blazhko effect results from the combination of two active modes of nearly the same frequency, one highly non-sinusoidal and one approximately sinusoidal.  That model was shown to accurately model results from RR Lyr, which is known to have a very asymmetric triplet spectrum, with the peak to the right of the fundamental being about 5 times the amplitude of the one on the left (measured ratio $R=0.194$ for Q2 \emph{Kepler} data).  In the current work, the inclusion of the lowest order interaction terms between the equations governing the two modes is shown to allow the adjustment of the side-peak ratio through the strength of the coupling, including ratios that are fairly close to 1.0 as is the case for V808 Cyg (measured ratio $R=0.857$ with main side-peak on the left of the fundamental).  In order to achieve this effect, the coupling strength must be below a critical value beyond which phase-locking occurs between the modes.  The adiabatic case produces a pure triplet spectrum, but when nonlinear damping is also included, it is found that additional odd-order side-peaks are generated.  This agrees with results from V808 Cyg which appears to favor the odd-order side-peaks over the even-order ones.  There appears to be no reason to expect a suppression of even-order side-peaks in other recent Blazhko models such as the ninth overtone model.  The concept of ``environment" is introduced, in which strongly excited modes can alter the frequency and amplitude of weakly excited modes by effectively altering the environment in which they reside.  This is a nonlinear effect and can be characterized by environmental terms (sometimes referred to as saturation terms) that appear in the equations of motion.  In the case of the Blazhko effect, the environment so produced is slowly varying at either the Blazhko frequency or at twice that frequency depending on the nature of the modes involved.  This varying environment leads to modulation of the weakly excited mode.  An example was found in the case of V808 Cyg for a mode which is apparently modulated at twice the Blazhko frequency.  This is shown to be consistent with the model, provided the two main modes are the fundamental and a nonradial mode with indices $m=0$ and odd $l$.  Other popular Blazhko models, such as the ninth overtone model, would appear to be unable to account for this doubling of the modulation frequency.

\appendix

\section{COMPLEX AMPLITUDE EQUATIONS}
\label{complex}

The complex amplitude equations may be obtained from the real equations as follows.  Start with the real set of equations governing the dynamics of a set of interacting modes:
\begin{equation}\label{realeqn}
\ddot{x}_j = - \omega_j^2 x_j + F_j(\textbf{x},\dot{\textbf{x}})
\end{equation}
where $x_j$ is the $j$th mode amplitude, $\omega_j$ is the corresponding frequency and $F_j$ is a general nonlinear function of all mode amplitudes and velocities [\citet{Nayfeh1994} also considers possible dependence on accelerations, but we do not discuss this here].  We introduce a complex variable $z_j$, the real part of which is the mode amplitude and the imaginary part is the negative of the mode velocity divided by the frequency:
\begin{equation}\label{zj}
z_j = x_j - i\dot{x}_j/\omega_j
\end{equation}
Taking the time derivative:
\begin{equation}\label{zjdot}
\dot{z}_j = \dot{x}_j - i\ddot{x}_j/\omega_j
\end{equation}
However,
\begin{equation}\label{xj}
 x_j = \mathrm{Re}(z_j) = (z_j+z_j^*)/2
\end{equation}
and
\begin{equation}\label{xjdot}
 \dot{x}_j = -\omega_j \mathrm{Im}(z_j) = i\omega_j(z_j-z_j^*)/2
\end{equation}
and from Equation~\ref{realeqn}
\begin{equation}\label{xjddot}
 \ddot{x}_j = -\omega_j^2(z_j+z_j^*)/2 + i\omega_jG_j
\end{equation}
where $G_j$ is just $-iF_j/\omega_j$ expressed in terms of the complex amplitudes and velocities, i.e. replacing $x_k$ with $(z_k+z_k^*)/2$ and $\dot{x}_k$ with $i\omega_k(z_k-z_k^*)/2$ for all $k$.
Thus from Equation~\ref{zjdot},
\begin{equation}\label{compeqn}
\dot{z}_j = i\omega_j z_j +G_j
\end{equation}
We should note that $F_j$ includes all additional terms appearing on the right hand side of Equation~\ref{realeqn} including linear and nonlinear damping and excitation for the $j$th mode as well as cross terms involving other modes.  So for example, if $F_j$ contains the linear excitation/damping term $2\kappa_j \dot{x}_j$ then $G_j$ will contain the terms $\kappa_jz_j$ and $-\kappa_jz_j^*$.  Note that $\kappa_j$ is the exponential growth rate for mode $j$ which combines both excitation and damping.  If negative, then damping exceeds excitation and the mode will decay exponentially to zero amplitude.  If positive then one or more nonlinear terms are needed to prevent the growth from continuing without bound.  If $\kappa_j$ is small compared to $\omega_j$, then it may be appropriate to ignore the term $-\kappa_jz_j^*$ as non-resonant since its dominant frequency is $-\omega_j$ instead of $\omega_j$.  In other words, if we can assume the $z_j$ is approximately proportional to $e^{i\omega_j t}$ then $z_j^*$ is approximately proportional to $e^{-i\omega_j t}$ which is strongly nonresonant.   A possible nonlinear damping term in $F_j$ would be $C\dot{x}_j^3$ with $C$ negative.  This expands to four terms in $G_j$, but the only resonant one is $(3/8)C\omega_j^2|z_j|^2z_j$.  The term $Cx_j^2\dot{x}_j$ produces a similar resonant term: $(1/8)C|z_j|^2z_j$.  There are two other cubic terms: $Cx_j\dot{x}_j^2$ and  $Cx_j^3$, both of which produce resonant terms that are purely imaginary functions times $z_j$ and so they will affect the phase but not the amplitude of the mode and they cannot provide damping.  None of the quadratic terms is resonant.

\section{FREQUENCY TABLES FOR V808 CYG}
\label{freqtab}

The data used from V808 Cyg is \emph{Kepler} project KIC 4484128, Q1 through Q17, long cadence corrected flux data.  The data was analyzed using Period04 software which attempts to fit the data to the form:
\begin{equation}\label{fit}
x(t)=Z+\sum_i A_i \sin(2\pi(f_i t + \phi_i))
\end{equation}
where $Z$ is the zero point (constant component of the flux), $f_i$ is the frequency, $A_i$ is the amplitude and $\phi_i$ is the phase for peak $i$ in the spectrum.  Note that the phase goes from 0 to 1 rather than $2\pi$.  The calculations are done in \emph{Kepler} flux units.  $Z$ for this data is 8835.47.  In Table~\ref{tabf0} is shown the fundamental mode $f_0$, its harmonics and associated side-peaks, while in Table~\ref{tabf2} is shown the mode $f_2$ and its side-peaks, which may be offset by multiples of the fundamental.  The two components of the split peak $f_2$ are labeled $f_{2a}$ and $f_{2b}$.  Frequencies in the two tables were calculated using $f_0=1.825273$, $f_B=0.01085$, $f_{2a}=2.697753$ and $f_{2b}=2.699816$.  Other frequencies in the tables are linear combinations of these.  They are assumed to be phase-locked and are therefore not allowed to independently vary from their calculated values.  Period04 was used to simultaneously determine the optimal amplitude and phase for all of these preset frequencies.  Reliability of these results was determined by examining the peaks in the spectrum and characterizing their visibility relative to the background noise.  Each peak gets one of three visibility ratings: good, which is used for peaks that are at least twice the height of nearby noise bumps in the data; low, which is used for peaks that are distinguishable from the noise bumps but are less than twice their level; and zero, which is used for peaks that are either not present or are not distinguishable from a noise bump.  The low cases could have a fairly high inaccuracy in the amplitude and phase determination.

\begin{deluxetable}{lllll}
\tablewidth{0pt}
\tablecolumns{5}
\tablecaption{Frequencies for V808 Cyg of the form $nf_0+mf_B$ for $1 \le n \le 4$ and $-5 \le m \le 5$\label{tabf0}}
\tablehead{ 
\colhead{Composition}&\colhead{Freq}&\colhead{Ampl}&\colhead{Phase}&\colhead{Vis}}
\startdata
$f_0-5f_B$&1.771023&8.79&0.975&Low\\
$f_0-4f_B$&1.781873&12.91&0.985&Low\\
$f_0-3f_B$&1.792723&49.27&0.190&Good\\
$f_0-2f_B$&1.803573&38.13&0.266&Good\\
$f_0-f_B$&1.814423&538.87&0.649&Good\\
$f_0$&1.825273&2857.07&0.364&Good\\
$f_0+f_B$&1.836123&463.26&0.811&Good\\
$f_0+2f_B$&1.846973&29.98&0.976&Good\\
$f_0+3f_B$&1.857823&56.07&0.281&Good\\
$f_0+4f_B$&1.868673&7.65&0.164&Low\\
$f_0+5f_B$&1.879523&5.66&0.632&Low\\
$2f_0-5f_B$&3.596296&5.73&0.313&Good\\
$2f_0-4f_B$&3.607146&3.75&0.791&Zero\\
$2f_0-3f_B$&3.617996&49.40&0.392&Good\\
$2f_0-2f_B$&3.628846&8.22&0.016&Good\\
$2f_0-f_B$&3.639696&449.92&0.815&Good\\
$2f_0$&3.650546&1507.69&0.599&Good\\
$2f_0+f_B$&3.661396&457.72&0.026&Good\\
$2f_0+2f_B$&3.672246&68.62&0.308&Good\\
$2f_0+3f_B$&3.683096&52.21&0.508&Good\\
$2f_0+4f_B$&3.693946&7.82&0.700&Low\\
$2f_0+5f_B$&3.704796&4.49&0.871&Zero\\
$3f_0-5f_B$&5.421569&4.24&0.417&Low\\
$3f_0-4f_B$&5.432419&5.82&0.806&Zero\\
$3f_0-3f_B$&5.443269&50.34&0.645&Good\\
$3f_0-2f_B$&5.454119&70.75&0.206&Good\\
$3f_0-f_B$&5.464969&421.69&0.048&Good\\
$3f_0$&5.475819&839.92&0.873&Good\\
$3f_0+f_B$&5.486669&391.00&0.299&Good\\
$3f_0+2f_B$&5.497519&85.09&0.654&Good\\
$3f_0+3f_B$&5.508369&49.78&0.761&Good\\
$3f_0+4f_B$&5.519219&12.62&0.077&Low\\
$3f_0+5f_B$&5.530069&2.74&0.213&Zero\\
$4f_0-5f_B$&7.246842&3.16&0.782&Low\\
$4f_0-4f_B$&7.257692&13.80&0.099&Good\\
$4f_0-3f_B$&7.268542&44.11&0.872&Good\\
$4f_0-2f_B$&7.279392&83.58&0.471&Good\\
$4f_0-f_B$&7.290242&348.07&0.349&Good\\
$4f_0$&7.301092&487.07&0.150&Good\\
$4f_0+f_B$&7.311942&312.60&0.583&Good\\
$4f_0+2f_B$&7.322792&96.64&0.953&Good\\
$4f_0+3f_B$&7.333642&37.60&0.090&Good\\
$4f_0+4f_B$&7.344492&14.87&0.374&Good\\
$4f_0+5f_B$&7.355342&3.15&0.654&Zero\\
\enddata
\end{deluxetable}

\begin{deluxetable}{lllll}
\tablewidth{0pt}
\tablecolumns{5}
\tablecaption{Frequencies for V808 Cyg of the form $nf_0+f_{2a}+mf_B$ or  $nf_0+f_{2b}+mf_B$ for $-1 \le n \le 2$ and $-6 \le m \le 6$ for even $m$\label{tabf2}}
\tablehead{ 
\colhead{Composition}&\colhead{Freq}&\colhead{Ampl}&\colhead{Phase}&\colhead{Vis}}
\startdata
$-f_0+f_{2a}-6f_B$&0.80738&1.05&0.936&Zero\\
$-f_0+f_{2b}-6f_B$&0.809443&4.12&0.009&Low\\
$-f_0+f_{2a}-4f_B$&0.82908&5.34&0.573&Good\\
$-f_0+f_{2b}-4f_B$&0.831143&6.40&0.871&Good\\
$-f_0+f_{2a}-2f_B$&0.85078&4.11&0.363&Low\\
$-f_0+f_{2b}-2f_B$&0.852843&3.25&0.192&Low\\
$-f_0+f_{2a}$&0.87248&22.64&0.725&Good\\
$-f_0+f_{2b}$&0.874543&16.50&0.264&Good\\
$-f_0+f_{2a}+2f_B$&0.89418&6.62&0.372&Low\\
$-f_0+f_{2b}+2f_B$&0.896243&8.25&0.882&Good\\
$-f_0+f_{2a}+4f_B$&0.91588&2.16&0.103&Zero\\
$-f_0+f_{2b}+4f_B$&0.917943&5.56&0.586&Low\\
$-f_0+f_{2a}+6f_B$&0.93758&2.51&0.343&Zero\\
$-f_0+f_{2b}+6f_B$&0.939643&5.69&0.883&Good\\
$f_{2a}-6f_B$&2.632653&3.37&0.856&Low\\
$f_{2b}-6f_B$&2.634716&9.09&0.042&Good\\
$f_{2a}-4f_B$&2.654353&11.63&0.600&Good\\
$f_{2b}-4f_B$&2.656416&12.44&0.886&Good\\
$f_{2a}-2f_B$&2.676053&10.03&0.389&Good\\
$f_{2b}-2f_B$&2.678116&3.50&0.115&Low\\
$f_{2a}$&2.697753&64.51&0.741&Good\\
$f_{2b}$&2.699816&46.00&0.267&Good\\
$f_{2a}+2f_B$&2.719453&15.46&0.396&Good\\
$f_{2b}+2f_B$&2.721516&22.57&0.933&Good\\
$f_{2a}+4f_B$&2.741153&13.22&0.172&Good\\
$f_{2b}+4f_B$&2.743216&15.64&0.632&Good\\
$f_{2a}+6f_B$&2.762853&1.04&0.658&Zero\\
$f_{2b}+6f_B$&2.764916&4.36&0.122&Low\\
$f_0+f_{2a}-6f_B$&4.457926&2.42&0.143&Zero\\
$f_0+f_{2b}-6f_B$&4.459989&6.14&0.316&Good\\
$f_0+f_{2a}-4f_B$&4.479626&7.66&0.865&Good\\
$f_0+f_{2b}-4f_B$&4.481689&9.61&0.182&Good\\
$f_0+f_{2a}-2f_B$&4.501326&10.61&0.676&Good\\
$f_0+f_{2b}-2f_B$&4.503389&5.32&0.388&Zero\\
$f_0+f_{2a}$&4.523026&39.98&0.994&Good\\
$f_0+f_{2b}$&4.525089&26.84&0.535&Good\\
$f_0+f_{2a}+2f_B$&4.544726&12.73&0.658&Good\\
$f_0+f_{2b}+2f_B$&4.546789&17.48&0.187&Good\\
$f_0+f_{2a}+4f_B$&4.566426&7.81&0.486&Good\\
$f_0+f_{2b}+4f_B$&4.568489&10.06&0.924&Good\\
$f_0+f_{2a}+6f_B$&4.588126&0.60&0.987&Zero\\
$f_0+f_{2b}+6f_B$&4.590189&3.28&0.311&Low\\
$2f_0+f_{2a}-6f_B$&6.283199&2.33&0.433&Low\\
$2f_0+f_{2b}-6f_B$&6.285262&5.68&0.565&Good\\
$2f_0+f_{2a}-4f_B$&6.304899&5.64&0.059&Good\\
$2f_0+f_{2b}-4f_B$&6.306962&6.64&0.401&Good\\
$2f_0+f_{2a}-2f_B$&6.326599&6.85&0.896&Good\\
$2f_0+f_{2b}-2f_B$&6.328662&3.28&0.676&Zero\\
$2f_0+f_{2a}$&6.348299&13.42&0.206&Good\\
$2f_0+f_{2b}$&6.350362&5.32&0.779&Good\\
$2f_0+f_{2a}+2f_B$&6.369999&8.94&0.872&Good\\
$2f_0+f_{2b}+2f_B$&6.372062&12.79&0.426&Good\\
$2f_0+f_{2a}+4f_B$&6.391699&5.14&0.697&Good\\
$2f_0+f_{2b}+4f_B$&6.393762&8.28&0.132&Good\\
$2f_0+f_{2a}+6f_B$&6.413399&0.72&0.316&Zero\\
$2f_0+f_{2b}+6f_B$&6.415462&2.60&0.587&Low\\
\enddata
\end{deluxetable}

\end{document}